# Technological Novelties of Ground-Based Very High Energy Gamma-Ray Astrophysics with the Imaging Atmospheric Cherenkov Telescopes

Razmik Mirzoyan

Max-Planck-Institute for Physics (Werner-Heisenberg-Institute)
80805 Munich, Germany, razmik.mirzoyan@mpp.mpg.de

**Abstract**

In the past three decades, the ground-based technique of imaging atmospheric Cherenkov telescopes has established itself as a powerful discipline in science. Approximately 250 sources of very high gamma rays of both galactic and extra-galactic origin have been discovered largely due to this technique. The study of these sources is providing clues to many basic questions in astrophysics, astro-particle physics, physics of cosmic rays and cosmology. The currently operational generation of telescopes offer a solid performance. Further improvements of this technique led to the next-generation large instrument known as the Cherenkov Telescope Array. In its final configuration, the sensitivity of CTA will be several times higher than that of the currently best instruments VERITAS, H.E.S.S., and MAGIC. This article is devoted to outlining the technological developments that shaped this technique and led to today's success.

## 1. Introduction

The classical book of Jelley [1] is a real jewel for researchers interested in Cherenkov radiation. It covers diverse aspects of the bluish emission and in great detail. Despite it being 65 years since the first edition of the book was introduced, it is still the table book of many researchers. Many highly regarded papers have been devoted to the history of Cherenkov emission and its use for ground-based very high energy (VHE) gamma astrophysics ([2-7]). For more details, the reader is invited to read the recent highly interesting book of D. Fegan [8], as well as the article from this author [9].

Some of the above publications show the chronological developments and a list of instruments built and operated in different countries. Unlike those publications, in this paper the author aims to highlight the chain of important technological developments, which improved the technique and allowed us to consider this branch of science as mature and established.

Below the author will to go into the details of important developments that led us to today's success.

# 2. Discovery of Cherenkov Emission in the Atmosphere

Galbraith and Jelley fixed a 25 cm diameter parabolic mirror of a short focal length inside a dustbin and set a 2-inch PMT in its focus, see Figure 1. In a series of experiments, they detected Cherenkov light flashes from air showers. Their discovery paper laid the foundation of the atmospheric Cherenkov light detection technique in 1953 [10]. Further developments of the latter gave rise to ground-based VHE gamma-ray astrophysics and led to today's powerful branch of astrophysics.

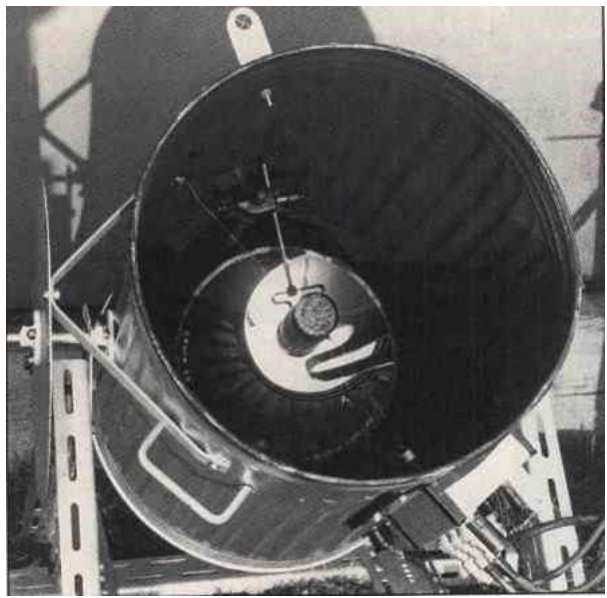

**Figure 1**. The original detector of W. Galbraith and J. V. Jelley described in [10].

## 2.1 First Generation Atmospheric Cherenkov Telescopes

Here we pose an interesting question. Would the early researchers working in ground-based VHE gamma-ray astronomy half a century ago have dreamed about its future scale and impact as a well-established branch of science?

Before proceeding with these important questions, we would like to prepare the reader with some basic information about the air showers, atmospheric Cherenkov light emission, the threshold of a telescope, and some other useful information.

2.1.1 Extensive Air Showers and the Cherenkov Light Emission

The earth's atmosphere is constantly bombarded by charged cosmic rays and neutral gamma photons. The charged particles and gammas, with energies in excess of several to a few tens of GeV, interact with the air molecules and trigger avalanche-like events known as extended air showers (EAS). To illustrate, let us imagine a 1 TeV gamma entering the atmosphere. In the vicinity of an air molecule, the incident photon can be converted into an electron-positron pair: γ → e− + e+. The latter has a very high energy and will therefore generate further gammas via the effect of bremsstrahlung in the electric field of the atomic nuclei. The above two-step cycle can be repeated in multiple rounds. Apparently, after each cycle, the number of secondary charged particles doubles while the original energy is shared among them. The height at which the number of secondary particles reaches the maximum is called the shower maximum. Typically, the secondary particles have an energy of ~300 MeV at the shower maximum. When the secondary particle energy decreases below the critical energy of ~84 MeV, the shower extinction phase starts.

A 2 TeV gamma can produce another generation of secondary particles compared to 1 TeV due to its twice higher energy. One can therefore assume that there should be a linear relationship between the incident energy and the number of secondary particles.

The latter move at a higher speed than light in the atmosphere; Note that this is possible because the refractive index of air is larger than in a vacuum, e.g., 1.00029 at sea level. Such particles produce Cherenkov light. The opening angle of the Cherenkov light cone θ can be calculated from the simple relationship cosθ = 1/n β, where n is the refractive index of air at the given altitude and β = v/c is the relative velocity of the particle (c is the speed of light). The 1 TeV gamma photon will produce ~130 Cherenkov photons per m2 in the wavelength range 300-600 nm, up to the so-called "hump" at ~130 m from the shower core (for an observation altitude of ~2 km above sea level). A typical electromagnetic shower has a time structure of 5-10 ns, as light and particles travel together (the latter being slightly faster) similar to a pancake that is a few meters thick.

This hump is the result of a kind of self-focusing effect of Cherenkov light production; the point of impact of a Cherenkov photon on the ground is the product of the Cherenkov angle with its production height. While the Cherenkov light emission angle continues to increase as particles penetrate deeper into the atmosphere (due to the increased density and refractive index of air), the product of the angle with height shows some focusing effect at 60-130 m from the shower core, see Figure 2 Left.

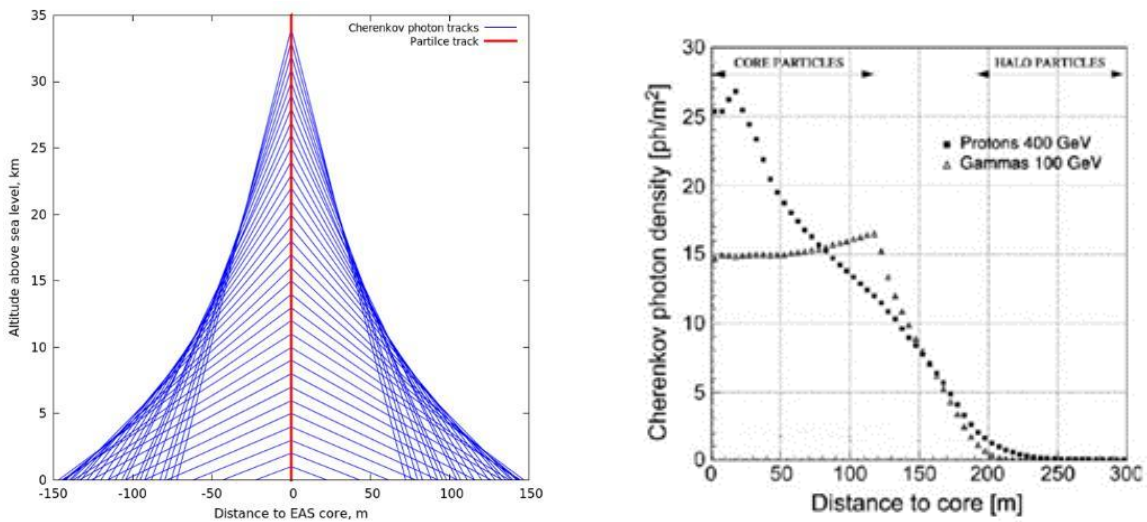

**Figure 2.** Left: lateral distribution of Cherenkov light emission from a single relativistic muon traversing the atmosphere. The observation height is 600 m a.s.l. Right: one can see the light focusing effect (the "hump") on the ground, at 70-130 m distance from the core. Image courtesy V. Samoliga. Right: The lateral distribution of Cherenkov light from a 100 GeV gamma and 400 GeV proton. One can see the "hump" at about 125 m distance from the shower axis (observation height 2.2 km a.s.l.). The "core particles" and the "halo particles" are meant to show the impact parameter range where the light arrives from high and low altitudes in the shower development, respectively.

During the development of an air shower, the e-e+ will emit Cherenkov light within a cone opening angles of ~0.2-1.2°. A 1 TeV gamma-ray shower is estimated to produce ~700 e- e+ pairs at the shower maximum. It may seem that the emitted Cherenkov light should be concentrated within a circle of a radius equal to the distance to the hump. In reality, the e- and e+ scatter multiple times along their motion, so the scattered light smears the light distribution on the ground and can form a large angle with respect to the shower axis, sending photons well beyond the hump. The multiple scattering angle is inversely proportional to the particle energy, i.e., the lower the energy, the larger the scattering angle. A simple estimate shows that the Cherenkov and multiple scattering angles (one sigma value) are about 0.6-0.7° for the e-e+ energy of ~1 GeV. For lower energies, i. H. for lower altitudes in the shower development, the multiple scattering effect becomes the dominant mechanism to spread Cherenkov light on the ground.

A hadron-initiated shower behaves similarly to that produced by a gamma, since in the hadron interaction, along with π+ and π-, π0 will also be born, which immediately decays into two gammas: π0 → 2γ. These two gammas will initiate electromagnetic cascades as described above. What makes the difference is that the charged π+ and π- also decay and the induced shower shows the hadron interaction signatures comprising hadrons, muons, neutrons, neutrinos, etc., superimposed on the electromagnetic showers. The differences between the hadron and electromagnetic showers will show up in their shapes and also in their lateral distribution of Cherenkov light density (see Figure 2b). One can see that, unlike protons, the gamma-rays can preferentially produce triggers in a measuring instrument for the impact parameter range ≤130 m.

The hadron showers will be much wider (due to the transverse momentum of the three-particle decay), longer, and with a chaotic structure, compared to gammas. Thus, a snapshot of both types of showers can help differentiating them. That is exactly what the contemporary imaging atmospheric Cherenkov telescope does; it can easily suppress hadrons by a few orders of magnitude while selecting gammas with high efficiency.

An imaging Cherenkov telescope of ~10 m2 mirror area will collect on average 130 ph/m2 × 10 $m^2$ = 1300 Cherenkov photons from a 1 TeV gamma-ray EAS. Assuming a ~10% conversion efficiency from Cherenkov photons to photoelectrons (ph.e.), one can expect to get 130 ph.e. for an image. Such an image can be well parameterized and used for efficient image selection.

Researchers have long wondered whether and how, for example, 50 GeV gamma-ray showers can be observed. Please note that such a shower provides only five Cherenkov photons per square meter area within the hump.

The formulation of this problem was due to the fact that many interesting phenomena were expected in the energy range below 300 GeV. In general, measuring the spectrum of a particular type of gamma-ray source over a potentially broad energy range and bridging it with the spectrum of lower-energy satellite missions could have provided a wealth of information.

Due to lower absorption by the extragalactic background light (EBL) the universe is becoming increasingly transparent to lower energy gamma rays, i.e., signals from distant active galactic nuclei (AGN), gamma ray bursts (GRB) and other possible remote transient events expected to become visible, see paragraph 6. Furthermore, the weak signals from pulsars were expected to become visible above the very low threshold ~10–20 GeV.

For a long time, up until the mid-1990s, the research community believed that measuring such low-energy events required the operation of expensive telescope facilities with unrealistically large mirror sizes, or converted solar power plants with several thousand square meters of mirror area. This is discussed in more detail later in Sections 6 and 7.

2.1.2 Chudakov's Telescopes in Crimea

Starting in 1960, A. Chudakov and colleagues built the first system of 12 telescopes with a total mirror area of 21 m2 in Crimea, near the shore of the Black Sea [11]. They intended to find out whether one can (relatively easily) measure a signal from some "prominent" celestial source candidates. These telescopes were simple parabolic searchlight mirrors of F/D = 0.6 m/1.55 m design with single 5 cm diameter PMTs in their foci. A diaphragm provided an aperture of 1.75° FWHM. To reduce the large aberrations and to improve the signal timing Chudakov designed a special lens set in front of the PMTs.

Each of the four independent mechanical mounts, set next to each other, carried three rigidly connected telescopes, see Figure 3. The pointing precision was 0.2° and 0.4° in elevation and azimuth, respectively. A coincidence system between these four telescopes had an advanced feature—a simple rate-stabilizing electronic circuit for counter-acting the variations of the light of the night sky (LoNS). The physical rate of showers was in the range of 3 Hz and the gamma-ray energy threshold was estimated to be ~3.4 TeV.

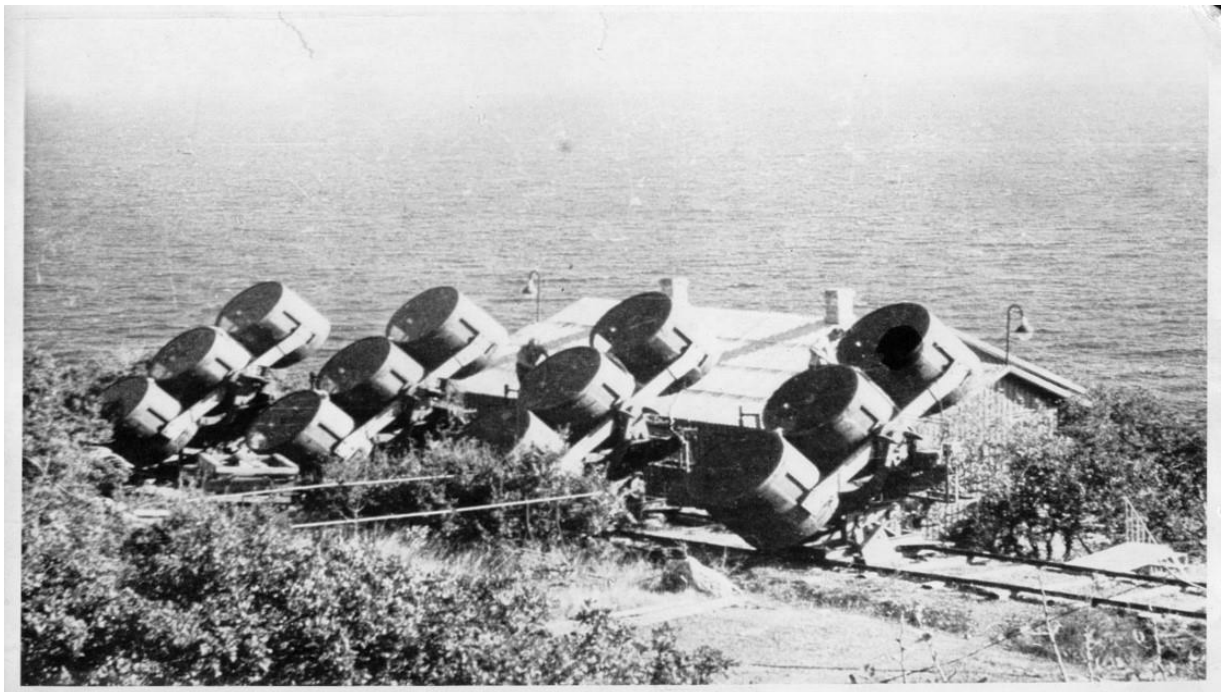

**Figure 3.** The telescope of A. Chudakov and colleagues in Crimea near the Black Sea shore.

Instead of continuous tracking, the so-called drift-scan mode was used for observations. These were performed by pointing in advance at the expected source position and waiting for it to slew through the field of view. Through repeated scans one could collect a reasonable amount of data during one night.

The list of observed sources is impressive given the limited knowledge of X-ray sources at the time. The educated guess was that radio sources should be good candidates to observe. They observed the Crab Nebula, Cygnus A, Cassiopea A, Virgo A, Perseus A, Sagittarius A. Moreover, the clusters of galaxies Ursa Majoris II, Corona Borealis, Bootes, and Coma Berenices were also observed.

The experiment was carried out for the duration of about 4 years. Unfortunately, they did not succeed in measuring a signal from any of the observed sources and, instead, derived only upper limits. For example, in the case of the Crab Nebula, the derived upper limit was about 20 times higher compared to its currently measured flux.

In principle, they could have discovered the gamma-ray emission from the Crab Nebula already ~60 years ago, but only at the expense of unreasonably long observations.

The list of sources above demonstrates that the researchers had a very smart observational program, even by today's standards.

It is not so uncommon in the history of science that a chain of incremental achievements step-by-step improve the technique and the technology, thus paving the road towards the goal to develop a new branch of science. Not surprisingly, this was initially performed by a handful of research groups scattered around the world. Perseverance may pay off and guide to the needed technology and technique.

At the ICRC conference in Moscow in 1959 G. Cocconi claimed that with a cosmic-ray instrument of ~1° resolution, operating at ~TeV energy range, one will measure a factor of one thousand times higher gamma-ray signal over the background from the direction of the Crab Nebula [12].

Unfortunately, this could not be confirmed by Chudakov and his crew and it had a sobering effect.

Interestingly, over a long run such overoptimistic claims turn out to be wrong. Nevertheless, they play an important role in sparking curiosity and generating activity in a particular field.

Chudakov's installation has probed the potential of the large size (~20 m2) array of the first generation non-imaging telescopes, operating in a narrow time coincidence. The hope of easy detection of cosmic gamma-ray sources turned out to be elusive, or at least more complex, than was originally anticipated.

A. Chudakov became a famous and influential researcher on the international level. Colleagues standing close to him used to tell that after non-detections of sources by his instrument, he became really skeptical about the prospects of the ground-based gamma-ray astronomy for a long time.

The distinguishing technical features of Chudakov's instrument were (a) the relatively large mirror area, which is essential for achieving a low detection threshold; (b) the narrow time coincidence between somewhat separated telescopes, aiming to counteract the effect of LoNS on signal detection; and (c) the best aperture for maximizing the signal/noise ratio by selecting a special diameter of a diaphragm.

## 2.1.3 Other First-Generation Telescopes

In the following years many smaller-scale experiments were built and operated. Several examples are listed below.

A new telescope of a very fast optical and electronic design, optimized for the pulsar studies, was built in Glencullen valley not far from Dublin in 1967-1968. It was based on four 0.9 m diameter F/2 mirrors (total area ~2.5 m2) and fast PMTs, put into coincidence with a gate width of 3.5 ns. The intention was to increase the signal-to-noise ratio by reducing the integrated charge from the LoNS. Later, that telescope was at first shifted to Harwell and then to Malta, where it started observations in early 1969. As a result, flux upper limits for several pulsars were set [13].

In another development, two 6.5 m diameter reflectors were set at a distance of ~120 m for stellar interferometry in Narrabri, Australia. In 1968, the researchers carried out observations of the Crab Nebula and two pulsars. No signal was measured [14].

Grindlay and colleagues made a step forward by using the "double beam" observation technique. Each telescope had two PMTs. While the main PMTs were inclined towards one another at 0.4° for observing the shower maximum region from a selected source direction, the other two PMTs were inclined to angles of 1.3° towards each other for measuring a signal from the so-called "muon core" of the showers. Though obscure and mysterious from today's point of view, the authors claimed that in this way they could halve the hadron background [15]. That was not much but the principle was interesting; upgrades and modifications of it will be widely used in the future. The author is still curious if one can imagine that effort as the first element of a form of two-pixel imaging?

The Haleakala telescope in Hawaii included six spherical, aluminum coated, coplanar glass mirrors of f/1 optics with 1.5 m focus, set on a single equatorial mount. Two independent sets of 18 PMTs in the focal planes observed separate areas of the sky, sharing the same set of mirrors. Each tube within a set collected light from a different segment of the total mirror area. The PMTs in the focal plane were operated in a fast single ph.e. detection mode [16]. When several tubes produced single ph.e. signals in a tight coincidence window, a trigger was produced. Later, it turned out that this detection technique suffered from exhaustive trigger rates from local muons.

The Nooitgedacht MK I telescope near Potchefstroom in the South African Republic consisted of four equatorially mounted mini telescopes (MT), set 55 m apart. An MT contained three light detectors, consisting of 1.5 m diameter, f/0.43 rhodium coated mirrors, focusing light on a XP2020Q PMT. Later, these were modified to the MK II telescope, which consisted of six MTs, set 225-322 m apart from each other. A single MT consisted of three mirrors, forming an f/1 optic with a focal length of 1.94 m. A PMT in its prime focus measured the ON source region, whereas the Cassegrain ring mirror in the focal plane around the telescope's axis reflected the light from the 4.5° OFF source region to a PMT installed on the mirror level. Thus, one could simultaneously measure the ON and OFF source regions. For details please see [17] and the references therein.

Another example is the first-generation telescopes of the Durham group, arranged similar to the logo sign of Mercedes [18]. Please note that due to the flatter "plateau" of the lateral distribution of Cherenkov light from gammas compared to hadrons, the separated by 50-

100 m distance arrays of telescopes could preferentially trigger on gammas at the hardware level (see Figure 2b).

The THEMISTOCLE array followed the approach of building a widely spaced array of 18 tracking telescopes in the south of France, each carrying a parabolic mirror 0.8 m in size. These provided a shower collection area in excess of 105 m2. This installation measured a signal from the Crab Nebula, but due to the small area of individual mirrors and the wide spacing, it had a high threshold of ~3 TeV and a low sensitivity [19].

The PACT experiment in India was of a similar design to THEMISTOCLE, but with higher sensitivity due to the larger size of both the mirror area of individual stations and the cluster of distributed stations [20].

The HAGAR telescopes [21] are located in Hanle in Himalaya, at 4500 m a.s.l., the same location as the 21 m diameter MACE imaging Cherenkov telescope [22]. HAGAR includes seven telescopes located at the center and corners of a hexagon inscribed in a circle of 50 m radius. The total reflector area of all the seven telescopes is about 31 m2. Each telescope consists of seven parabolic glass mirrors of 0.9 m diameter. Results on the Crab Nebula detection were recently published by this telescope [21].

The AIROBICC instrument of HEGRA on the Canary island of La Palma operated for almost ten years, starting in 1992. It comprised an array of 100 optical detector stations, each based on an 8-inch size PMT, coupled to a Winston cone-type light concentrator. These stations were placed next to the particle (scintillator) detector array of HEGRA for simultaneously measuring the Cherenkov light to particle density from air showers. AIROBICC stations measured integrated Cherenkov light in a wide field of view of ~1 sr. The fast timing between the detector stations allowed for the measurement of the incoming direction of showers with a high precision [23]. Due to the LoNS integration in a wide field of view, the threshold of AIROBICC for gamma rays was estimated to be a few tens of TeV. The relatively small size of the array, coupled to the high threshold, did not enable significant gamma-ray source detections.

Except for the 10 m diameter Whipple telescope, which played a central role in giving birth to gamma astronomy (this will be discussed in some detail below), no major technical improvements were achieved until the 1980s. In Figure 4, one can see a photo of the 10 m diameter Whipple telescope.

As a rule, researchers used 0.6–1.5 m diameter military searchlight mirrors of the parabolic shape of F/0.5 optics that suffered from poor angular resolution. They used coincidence between several such mirror elements, which enabled the lowering of the energy threshold of the instrument. Further, some of them used a number of PMTs for simultaneously monitoring the source and the background regions.

Most of these are reflected in the proceedings of the workshop series of "Towards a Major Atmospheric Cherenkov Detector" as well as in proceedings of the international cosmic ray conferences.

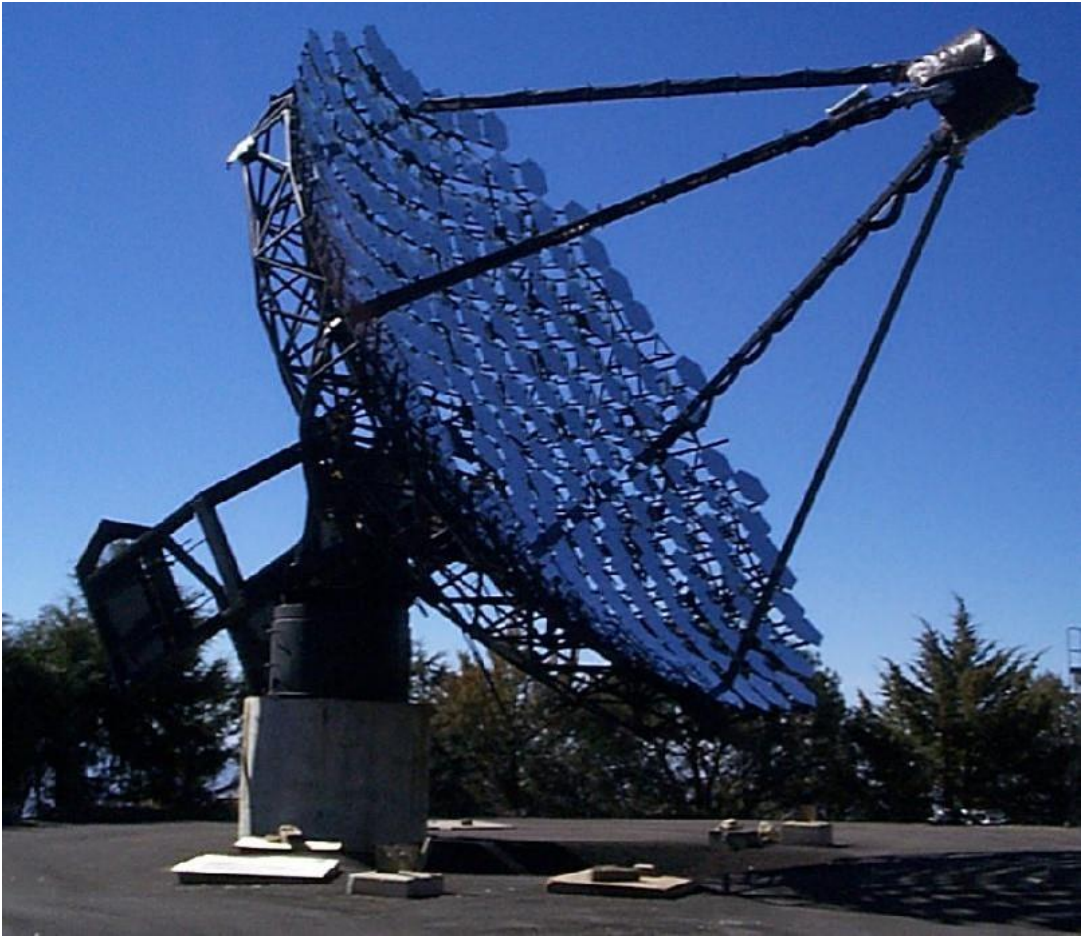

**Figure 4.** Photo of the 10 m diameter pioneering Whipple IACT on Mount Hopkins in Arizona.

2.1.4. A Short Summary on the First-Generation Telescopes

Please note that most of those experiments were based on counting the number of excess events from the ON-source and selected OFF-source regions. The fluctuations of the LoNS play an important role because its instantaneous positive fluctuation can add-up with a genuine small signal from a given shower and produce a trigger. Because of the natural differences of the LoNS intensity in ON and OFF-source regions, such brightness differences can produce a positive or negative excess.

Early on researchers learned to counteract the brightness differences in ON and OFF regions by inserting a weak light source next to the PMT so that the sum light from it and LoNS stayed constant to a few percent precision [11]. However, whether the achieved precision was enough for long observations still remains a sensitive question.

One should keep in mind that these were (are) non-imaging telescopes, the output signal of which solely depends on the count rate through the set-in front of the PMT diaphragm.

The technique used in the first-generation telescopes has improved over time. The researchers understood the main issues and made important developments, which paved the road for the next generation imaging telescopes.

The majority of the first-generation telescope installations from time to time reported "signals" measured from some sources, including pulsars. While one could imagine that sometimes some unknown at the time upper atmosphere light phenomena may have produced a short, sporadic excess, in the case of pulsars one of the probable reasons was the improper statistical and systematic treatment of the data (see, for example, [15]).

Unfortunately, those reports could not stand any serious criticism from today's point of view. Those can be understood by considering the strong desire of the small, enthusiastic groups of researchers to over-interpret small, insignificant and/or sporadic excess from observations as indications of the searched for signal from sources.

It seems that over time, the unwritten rule of "source code" (analogous to "dress code") encouraged researchers reporting source detections at conferences.

Such source reports culminated by the end of 1980s (see [24]).

The net impact of these questionable reports proved important, as it allowed the discipline to be kept alive in its infancy.

## 3. EAS Images in Cherenkov Light Obtained Using an Image Intensifier

The measurement of air shower image shapes by using an image intensifier by Hill and Porter in 1960 [25] can be considered as a real milestone. They coupled a 25 cm diameter wide-angle Schmidt telescope to an image intensifier and photographed the images of air showers. At some point they understood the potentialities of ground-based gamma-ray astronomy. At first, they noticed that the elliptical-shape shower images were offset from the source direction as well as understood that the image shape depends on the impact point of the shower axis. From here it was a stone's throw to the idea that by using two separated by some distance telescopes, one can derive the incoming direction of parent particle as well as largely suppress the background.

This important aspect has been demonstrated by the stereoscopic systems of Imaging Air Cherenkov Telescopes some 30 years later. It was the HEGRA collaboration that demonstrated the advantages of the so-called "stereo" observations, see more below. Further, the Crimean GT-48 telescopes pursued a similar goal, but their results remained more than modest due to the separation of telescopes by only ~20m. Because of this, the telescopes measured nearly the same images and the image parameters were strongly correlated, see more below.

In the summer of 1985, during one of the visits to Crimea, the author asked the GT48 group leader Arnold Stepanian why he put the telescopes so close to each other. He answered: "show me a single installation that has a higher count rate of air showers than mine". Thus, this complex array served to also provide a low detection threshold, which incidentally was ~0.9 TeV according to their simulations.

# 4. The First Monte Carlo Simulations and the "Stereo" Observations

Victor Zatsepin (not to be confused with Georgy Zatsepin from GZK cutoff), a crew member of A. Chudakov, published a remarkable Monte Carlo study paper in 1964 [26]. He obtained the equal photon density contours of air shower images produced by gamma rays as well as their angular distribution and radial photon densities. It is striking to read in that paper *"since the maximum intensity of the light from a shower does not coincide with the direction of arrival of the primary particle, in researches in which the determination of the angular coordinates of the primary particle is made by photographing the light flash from the shower one should seek improved accuracy in this determination by photographing the shower simultaneously from several positions"*.

From the above it can be seen that some researchers clearly understood the potential of coincidence measurements, now better known as stereoscopy, some 60 years ago.

The experiments with image intensifiers continued for some more years, but no further breakthrough occurred.

# 5. The Second-Generation Telescopes

## 5.1. The 10 m Whipple Telescope

In 1967, Giovanni Fazio and colleagues began constructing a 10 m diameter, F/0.7 telescope on Mount Hopkins, at the Whipple observatory, at a height of 2300 m a.s.l. [27]. The large diameter of the reflector of the telescope, together with a fast PMT in the focal plane, provided a relatively low detection threshold for air showers. The telescope started operating in 1968, initially with a single 5-inch PMT in the focus. Afterwards the number of PMTs was at first increased to two and later to ten PMTs for simultaneous ON and OFF source observations. In 1968 Trevor Weekes co-authored a publication with G. Fazio and two more colleagues about observations of 13 gamma-ray source candidates. The Crab Nebula was prominently in the list, but M87, M82, IC443 were also observed [28]. Only flux upper limits above the threshold of ~2 TeV were derived. As we know from later observations, the listed sources turned out to be gamma-ray emitters.

Already in 1977, T. Weekes and E. Turver suggested using a system of two telescopes separated by 100 m, equipped with 37-pixel imaging cameras. The intention was to strongly suppress the background [29]. The first imaging camera used 37 pixels of 0.5° size in a hexagonal configuration and covered a field of view of 3.0° in the sky. This camera was installed on the 10 m telescope in ~1983.

The next key development was the suggestion of Michael Hillas to parameterize the images by using the second moments of the measured charge distributions in the camera plane [30]. Interestingly, it is one of the rare cases where a conference contribution paper collected a huge number of citations. Using this formalism, the Whipple team succeeded in measuring the famous 9 sigma signal from the Crab Nebula in 1989 [31].

This is considered as the birthday of the ground-based VHE gamma astronomy. The scientific intuition and perseverance of Trevor Weekes and the small team around him paid off after ~20 years of effort and gave birth to a new branch of science.

The technological novelties of the Whipple telescope were the use of the Davies-Cotton optical design [32], for the 10 m diameter reflector and the 37-pixel imaging camera in its focus. A few years later this camera was exchanged for a finer resolution one, employing pixels of 0.25° in size. This has significantly improved the telescope's sensitivity and allowed to lower its threshold from 700 GeV down to ~300 GeV.

## 5.2. GT-48 in Crimea

Since the late 1960s the group in the Crimean Astrophysical Laboratory (CrAO) led by Arnold Stepanian used two parabolic searchlight mirrors of 1.5 m diameter in coincidence for studying gamma sources. They reported detections of Cassiopea and Cyg X3 in the early 1970s, with the latter getting a particularly strong response from the community. In the 1980s, the group started constructing a set of two large telescopes, separated by 20 m distance, named GT-48. On each mount they built six telescopes, three of the imaging type with 37 pixels and another three operating a single UV-sensitive, solar blind PMT. Every telescope had four mirrors of 1.2 m diameter and 5 m focus. The goal of the Crimean group was to profit from the stereo observations, see, for example [33]. Because they did not want to sacrifice neither the threshold nor the coincidence rate, they put the telescopes at 20 m distance from each other. Their relatively small reflectors and the low altitude of the location of 600 m a.s.l. provided a threshold of 900 GeV. The proximity of the telescopes did not allow them to fully exploit the differences in image parameters otherwise seen from largely separated detectors.

In 1989, this installation was put into operation and in subsequent years it measured a number of sources.

The technological novelties of the Crimean GT48 were the two sets of telescopes separated by 20 m, the used solar blind PMTs for measuring the UV content of air showers (the idea was that the muons in hadron showers produce more UV light) and the use of the coincident technique.

## 5.3. High Energy Gamma Ray Astronomy (HEGRA)

The first telescope of HEGRA was designed in 1990, as a somewhat modified version of the Yerevan Physics Institute (YerPhI) first Cherenkov telescope [34]. The latter was the prototype of the planned five telescope "stereo" array (proposal from 1985) which was later adopted by the HEGRA collaboration. Further developments of it became known in the community as the HEGRA air Cherenkov telescopes [35].

Originally each telescope was planned to have a 3 m diameter tessellated mirror of 5 m$^2$ area and to be equipped with a 37-pixel imaging camera in the 5 m focal plane. The pixels used light guides of a conical form (focons), made of UV transparent Plexiglas and subtending an angular aperture of 0.41° [36]. The imaging camera was based on the Soviet FEU-130 type special PMTs with GaP first dynode, providing a gain of 25–30 and thus a very high amplitude resolution. The very high-quality glass mirrors were produced in Yerevan Physics Institute. The mechanical mount of the first telescope was installed on the Roque de los Muchachos observatory in La Palma in late fall 1991 and the camera in mid-1992.

A ~5 sigma hint of the first signal from Crab appeared after two months of data taking, in late fall 1992 [37]. In the following year the second telescope with the same pixel size but with one more ring in the camera (61 pixels) and a larger reflector of 4.2 m was built and put into operation at ~100 m distance from the first one. The stereo observations, the power of which had been predicted in a dedicated Monte Carlo study paper in 1993 [38], could start.

In the following years, four more telescopes of the same size as the second one but with cameras of 271 pixels with a size of 0.25° were installed. In the end, the second telescope, too, was given a 271-pixel camera and the array was completed in 1997. The last upgrade in the same year was the increase in the mirror area of the first telescope to 10.3 $m^2$. In Figure 5 one can see a photo of the HEGRA array.

HEGRA operated until 2002. It convincingly demonstrated the long-awaited power of stereo observations [39] and produced a wealth of scientific results.

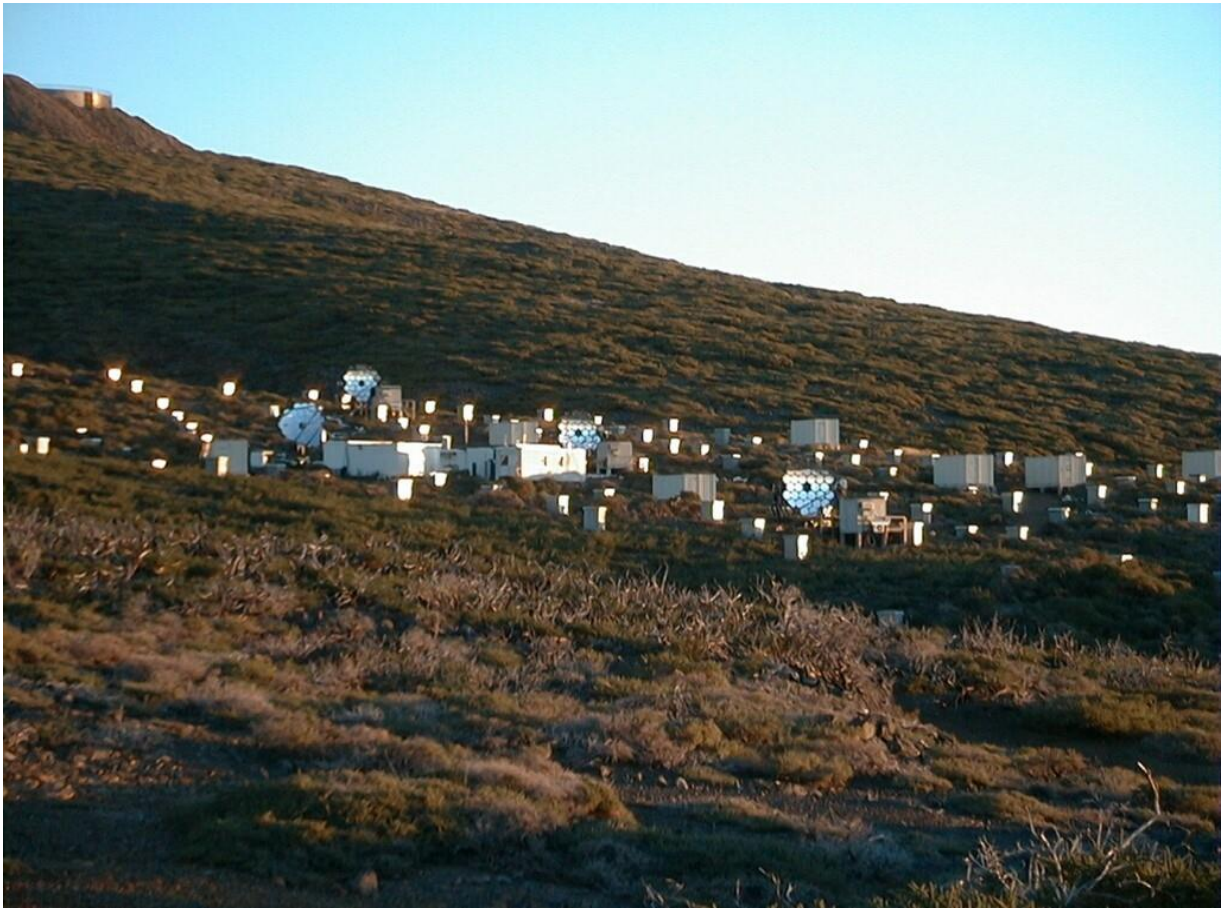

**Figure 5**. Photo of the HEGRA array. One can see four out of the six IACTs of HEGRA.

These second-generation imaging telescopes provided only a handful of sources, but it became clear that still there was a big potential in the "stereo" technique that was just waiting to be explored more extensively.

## 5.4. The 7-Telescope Array

The Japanese 7-Telescope Array was originally planned as a detector that included two arrays, each with 127 imaging telescopes, operating in coincidence [40]. Each telescope had a 3 m diameter mirror and a 256-pixel camera. In 1996–1997, three out of seven such telescopes were built and installed in Dugway proving grounds, Utah, USA. The remaining four were planned to be installed within one year. The telescopes started taking data on several interesting objects as, for example, the flaring MKN-501 and 1ES1959.

Unfortunately, a ~6 m long unarmed military missile lost its target and instead hit and destroyed the data taking containers of the experiment. Thus this array operated for less than one year in 1997.

The square PMTs and light guides used in these telescopes were innovative.

## 5.5. CLUE

The Italian CLUE collaboration tried to extend the application range of the IACT technique into the deep near UV range. They installed an array of nine 1.8 m telescopes at the HEGRA site on La Palma. They used a multi-wire, UV-light-sensitive proportional chamber (MWPC) for recording the EAS images. A matrix of electrodes in the rear side of the camera allowed to read out the images. The imaging camera was filled with a gas mixture containing TMAE. It was believed to provide a quantum efficiency of 5–15% in the range 190–230 nm. TMAE turned out to be an aggressive substance, attacking the camera materials. Further, the short distance of Cherenkov light absorption in the chosen wavelength range limited the imaging capabilities. These telescopes operated in 1997–1999.

Clue reported detections of the Crab Nebula, Mkn 421 and Mkn 501, and the lunar shadow [41].

## 5.6. CAT

The French CAT telescope, put into operation in late 1996 [42], on the same site as the previous non-imaging ASGAT [43] and THEMISTOCLE [19] instruments, started operating a 600-pixel high-resolution imaging camera based on a pixel size of 0.12°. Soon after the telescope was put into operation, the researchers found out that due to the very fast pulses from the PMTs, the detection efficiency of gamma-rays was quite low. After slowing down the speed of pulses towards ≥2.5 ns, they recovered the high efficiency for triggering gamma-rays. To counteract the bending of the relatively fragile mechanical frame of the telescope, they used data from the several imaging cameras installed on the structure. This was a successful telescope, which provided very interesting results.

## 5.7. CANGAROO

CANGAROO was a collaboration between several universities from Japan and the university of Adelaide. The collaboration started operating a 3.8 m size single telescope of parabolic shape that had been used for lunar ranging in the past. It started operating in 1992 at a threshold of a few TeV and in the following years discovered several new sources of gamma rays. Ten years later, four telescopes of 10 m size were built. These telescopes made a number of discoveries and very interesting observations. The telescopes had some differences in the design. Along with technical problems, mostly related to the chosen type of mirrors, there were also technical and organizational problems related to the data analysis. When the H.E.S.S. telescopes started to become operational in 2002–2004, they could not confirm some of the CANGAROO results [44]. A few years later this array terminated its operation.

# 6. The Very Low Threshold, EBL and Solar Power Plants as Gamma-Ray Telescopes

It was recognized rather early that mirror-based solar power plants can offer large mirror areas of several thousand m2 that could be used for collecting scarce photons from sub-100 GeV gamma showers. There was no measuring instrument and the unexplored energy range 10–300 GeV was considered as "terra incognita". Many interesting physical phenomena were expected there. The universe is full of photons emitted by galaxies and stars during its evolution. The EBL photons could be thought of as a kind of "gas" filling the space. The complex spectrum of this light extends from UV to far infrared, see for example [45].

When a very high-energy gamma ray travels through space from a cosmologically distant source, it can interact with one of these low-energy photons. If the energy of these two photons in the center of mass system is more than twice the rest mass of the electron, an electron–positron pair can be produced. This is an energy dependent phenomenon that limits the visibility of gamma ray sources in the universe; the higher the energy, the stronger the absorption. This weakens the flux of gamma rays from distant sources. The situation changes dramatically when moving towards the energy range below 100 GeV, down to ~10–20 GeV; the universe is becoming more and more transparent and very distant sources could be observed. Just to give the reader a feel for it; from the famous Mrk-421 and Mrk-501 sources, located at the redshift of ~0.03, the measured highest energy photons are limited to below 20 TeV due to strong absorption. However, if the source is located at the redshift of ~1, given a strong signal, some photons with energies up to ~200 GeV could still survive. Signals from pulsars, from distant AGN, from GRB and from various transient events were anticipated in the unexplored energy range 10–300 GeV.

The researchers were discussing about the possibilities if and how one can lower the threshold of a Cherenkov telescope by more than one order of magnitude. In the beginning of 1990s the threshold energy of the 10 m Whipple telescope of ~75 m$^2$ reflecting area was estimated to be 300–400 GeV. The common belief was that for lowering the threshold energy of a telescope by a factor of $n$ one needs to increase its mirror area by $n^2$ times. So, for example, for lowering the threshold energy of ~1 TeV of the ~10 m$^2$ HEGRA CT1 telescope by a factor of 20 down to ~50 GeV one needed to increase its mirror area by 400 times, i.e., one needed a mirror area of 4000 m$^2$!

For the very low threshold project, it was proposed to build an array of nine telescopes, each 100 m in diameter [46]. Compared to the latter option, the existing solar power plants with their distributed mirror surface area of several thousand square meters seemed to offer an interesting alternative. Several solar power plants were rendered into gamma-ray detectors. The technique of doing that was quite different for individual research teams from STACEE (NM, USA), CACTUS (CA, USA), CELESTE (France) and GRAAL (Germany and Spain). For example, while the GRAAL team [47] was attempting to collect Cherenkov photons from heliostats in the field into a ~1 m-size Winston cone, STACEE tried to organize a kind of imaging in the central light collection tower, directing light from individual heliostats to specific PMT channels [48]. For a comprehensive review of converted solar power plants please see [49].

Some interesting measurements were performed by using these arrays. The French CELESTE instrument tried to measure flux from the Crab Nebula down to ~60 GeV [50]. The comparison with today's precise measurements show that their reported flux was 2.5 times too low.

With the operation of the MAGIC telescope one confirmed that the above assumed relation of the threshold on the mirror area was wrong. As predicted, the threshold was inversely proportional to the mirror area [51,52], i.e., for reducing the cited above CT1 threshold from 1 TeV down to 50 GeV, one needed to increase the mirror area only by ~20 times, i.e., to build a telescope with ~200 m$^2$ mirror area (see the next paragraph for more details).

It became obvious that the "classical" imaging method was able to provide much higher efficiency than the solar power plant detectors, so shortly afterwards these ceased their operation.

## 7. The Threshold of an Imaging Air Cherenkov Telescope

It is interesting to note that the MAGIC-I telescope, that has only 236 m$^2$ of mirror area, could successfully perform measurements also in the sub-100 GeV energy range, down to 50 GeV. This is in striking contrast to the above-mentioned statement about the threshold dependency on the mirror area.

The issue of the threshold is interesting to illustrate on the example from a publication by K. E. Turver and T. C. Weekes from 1981 [18]. There one can read:

*"The energy threshold of a simple detector is inversely proportional to the diameter of the light collector. An energy threshold of $10^{11}$ eV requires an effective aperture of 5–10 m. To reach $10^{10}$ eV requires an aperture of 50–100 m; such apertures would have been out of the question a few years ago but the development of large concentrators for solar energy research makes this energy threshold a realistic possibility"*.

The above-mentioned dependence of the threshold on the mirror area, that even today is circulating in some publications, is not correct. Please note that for sub-100 GeV gamma-ray astronomy the authors refer to the use of large concentrators for solar energy, which in fact some 20 years later has happened (see Section 6).

Unlike the non-imaging detectors, the lower threshold of an imaging telescope is simply inversely proportional to the used mirror area (or to the squared-diameter). This can be explained by the fact that for an imaging telescope it is not the fluctuations of the LoNS that set the lower threshold, because the LoNS in the field of view is "split" between a large number of pixels, which in addition are put into some coincidence scheme.

A higher-level requirement is that for analyzing an image one needs some minimum amount of charge, on the order of ~100 photo electrons [51]. Realization of the latter relation played a key role for enabling the successful operation of the IACT technique in sub-100 GeV energy range, down to 10–20 GeV. This was substantiated by proposing and building the pioneering 17 m diameter MAGIC telescope project for sub-100 GeV gamma-ray astrophysics [52].

## 8. The Third Generation Telescopes

The third-generation telescopes were designed before the potential of the second-generation telescopes was fully exploited. Already, in 1995, the first presentations on the concrete concept of 17 m diameter MAGIC were made [53,54]. These were followed by the VERITAS letter of intent in fall 1996 and in the next year by H.E.S.S. Both VERITAS and H.E.S.S. were following the goal of conducting astrophysics with a stereo system of 10 m diameter telescopes, based on well-known, proven technologies. These were well-known thanks to the Whipple telescope and the fresh experience of HEGRA. Instead, the design of MAGIC aimed to move to the sub-100 GeV energy range, down to 20–30 GeV, into the "terra incognita". Obviously, this task was significantly more demanding and challenging, and several novel techniques and technologies were necessary for making it possible.

When HEGRA stopped operating in 2002, the collaboration split into two parts. One part together with the scientists from France, largely people from the CAT experiment, made the core of the H.E.S.S. collaboration and built their instrument in Namibia. The other part stayed in La Palma, at the original site of HEGRA, and together with scientists from Spain and Italy founded the MAGIC collaboration.

### 8.1. H.E.S.S.

The application of the H.E.S.S. collaboration was supported by the German and French financial agencies (while the VERITAS team had to wait for several more years to secure the financial support). The H.E.S.S. collaboration built their telescopes and started operation in Namibia in 2002–2004. At the beginning, the H.E.S.S. team performed a scan along the galactic plane and developed a really rich harvest of galactic sources. This array has turned out to be a very successful instrument, making a really high number of important discoveries and measurements above the energies 160–200 GeV, see, for example, [55].

A very large telescope of 28 m diameter was set in the center of H.E.S.S. in 2012. This also allowed them to perform observations in the very low energy range of a several tens of GeV [56]. The design and construction of the H.E.S.S. telescopes followed a conservative approach, based on proven technologies.

## 8.2. VERITAS

The VERITAS telescopes, unlike H.E.S.S., who operate imaging cameras of ~5° geometrical apertures, use cameras of 3.5° field of view. Otherwise, both instruments are similar, and both have increased the originally planned 10 m diameter of their telescopes to 12 m. For some time, the exact location of these telescopes in Arizona remained uncertain. In the end VERITAS was built next to the administrative building of the Harvard-Smithsonian Center for Astrophysics, not far from the foot of Mount Hopkins and was inaugurated in 2007. As one would expect, VERITAS also turned out to be a very successful instrument that, in recent years, has realized a high number of important discoveries and measurements, see, for example [57].
The design of the Veritas telescopes, similar with H.E.S.S., followed a conservative approach.
One should mention the upgrade of VERITAS with high QE bialkali PMTs in 2012 [58], which lowered the threshold down to ~90 GeV.

## 8.3. MAGIC

In the mid-1990s, an energy scale below ~300 GeV, down to 10-20 GeV, was considered as "Terra Incognita", simply because there was no instrument, neither on ground nor in space, to observe.
The intention to build MAGIC was to operate a ground-based instrument in the energy domain of ≤300 GeV, down to ~10 GeV. In the mid-1990s that was considered impossible. Initially, mostly because of the financial and organizational constraints, the 17 m diameter MAGIC was proposed as a stand-alone telescope [52-54]. Several innovations were necessary for operating the single telescope in the very low energy range of <100 GeV, where a strong background from local muons was expected. MAGIC researchers hoped to strongly suppress the different backgrounds by means of an ultra-fast opto-electronic design of the telescope. For this purpose, a reflector of parabolic design was chosen, which can provide, for example, time resolution of ≤140 ps within the 1° field of view. Spherical shape mirrors of 11 different radii of curvature, laid on the reflector, provided a good approximation of the intended parabola. Along with this, very fast hemispherical PMTs were developed for the needs of MAGIC by the company Electron Tubes from England. In combination with the light guides and a mat lacquer coating, these provided an enhanced quantum efficiency [59].
The PMT analog signals were converted into light by using Vertical Cavity Surface Emitting Laser (VCSEL) diodes operating at ~850 nm. This light, by using optical fibers, was transported to the electronic room, where it was converted back into electrical pulses with practically no degradation of time features.
The MAGIC-I telescope was built and put into operation in 2003-2004. The fast signals were initially read out by using 300 MSample/s custom-built FADCs.
Starting in 2007 MAGIC-I used 2 GSample/s fast multiplexed FACDs for the read-out [60].

The measurements showed a bandwidth of ~230 MHz for signal channels. This ultra-fast timing allowed MAGIC to suppress further down the contribution from the LoNS as well as the hadron-induced background by a factor of 2–3.

For the first time one could significantly enhance the sensitivity of a detector based on fast timing [61]. A single telescope, which measures a given shower projected on its imaging camera as a two-dimensional, flat image, due to fast timing (one image every 500 ps) can also measure data about the third, perpendicular to the camera direction, i.e., it can scan the image in three directions, coming close to stereoscopic imaging.

By developing a special so-called SUM-trigger configuration (see more below), the researchers could operate the stand-alone MAGIC-I telescope even at a very low threshold of ≥25 GeV. This allowed them to discover a pulsed signal from the Crab pulsar, which made a strong impact on the pulsar theory models [62].

The next serious improvement of MAGIC's sensitivity was due to the construction and operation of an almost clone telescope, at 85 m distance from the first one in 2009, see Figure 6. This has essentially doubled the sensitivity of the first telescope. By using the standard trigger, MAGIC performed observations of some selected sources as, for example, the Crab Nebula and its pulsar, at energies as low as ~50 GeV [63].

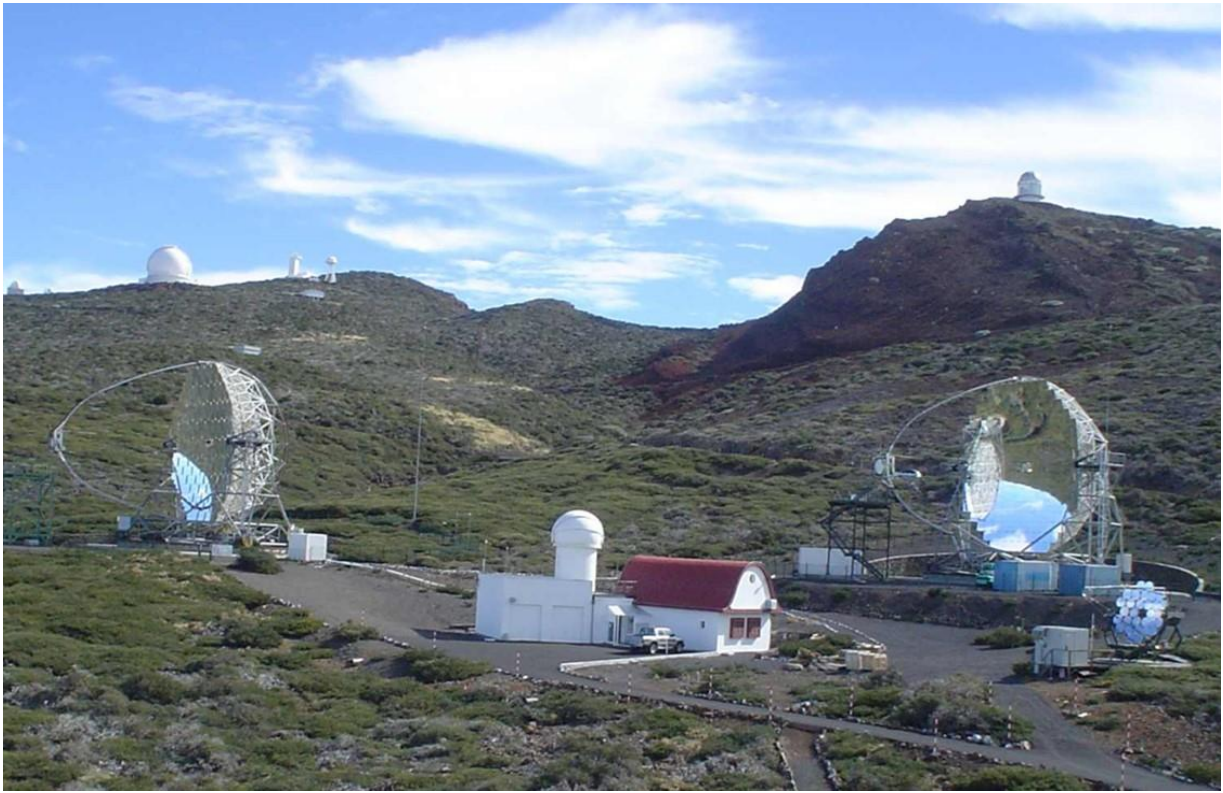

**Figure 6**. The two 17 m diameter MAGIC IACTs at the ORM observatory in La Palma. In the center one can see the experimental house. The white dome on top of the garages harbors the LIDAR instrument. In the top left corner one can see the famous 4.3 m diameter William Herschel telescope. The 2.5 m diameter Nordic optical telescope is located on the top-right summit. In the lower right corner one can see the 4 m FACT telescope.

The imaging cameras on both MAGIC-I and MAGIC-II were upgraded in 2012. Along with novel, higher QE hemispherical PMTs from Hamamatsu,

developed for MAGIC, a new capacitive memory based FADC readout system on DRS4, operated at ~2 GSample/s, was introduced.

The MAGIC telescopes introduced a number of novelties into the field, some of which later became the standard.

While some of the novelties were important from the operation reliability point of view (fully sealed, actively temperature controlled and stabilized imaging camera by circulating a liquid coolant in a closed loop), others (analog signal light transmission via optical fibers for preserving the fast speed of pulses, hemispherical input window PMTs, coupled to tailor to these light guides of special design, intended to enhance the detection efficiency of photons due to double crossing of the PMT photo cathode by the impinging light) helped to further improve the technique.

Special attention was paid to produce a light-weight reflector frame from reinforced carbon fiber, for the reduced weight of the telescope and low needed momentum for fast repositioning. This was considered as an important step for promptly reacting to alerts from satellite missions on transient sources, such as, for example, GRBs.

Though delayed, nevertheless this feature has fully paid off on January 2019, when for the first time the most intense gamma-ray signal at VHE was measured from the A 190114C only one minute after its explosion [64,65].

Of course, the light-weight reflector frame was not for free; it bends under varying gravitational loads when tracking a source. To counteract the deformations, an Active Mirror Control system was developed. While tracking a source, this adjusts the direction of every single mirror of ~1 $m^2$ area under computer control, providing the best optical point spread function (PSF) in the focus [66].

The other novelty was related to ultrafast speed of 2 GSample/s readout of the data. This very fast readout of the data at every 500 ps allows one to obtain multiple images from one and the same shower, tracing its development in time. Moreover, by integrating the signal charge in only ~3 ns time window, one effectively suppresses the LoNS contribution.

All these has enhanced the background rejection power, allowing for the operation of the telescopes in the energy domain of ≥20 GeV (see below).

8.3.1. SUM Trigger for MAGICs

One of the main obstacles for obtaining a low-threshold setting for an IACT is the adverse effect of after-pulsing in PMTs [67].

The standard trigger threshold of the two MAGIC telescopes was halved by using a so-called Sum-Trigger. Along with a circuit to suppress the importance of the after-pulsing, the Sum-Trigger detects weak, loose images in ~0.5° wide patches. As already mentioned, in 2008 it allowed for the revelation of pulsations from the Crab pulsar for energies of ≥25 GeV [61].

The novel, professional version of the Sum-Trigger-II, developed for the two MAGICs to work in coincidence, recently allowed the detection of a very weak signal from the Geminga pulsar at energies of ≥15 GeV [68].

Physics novelty introduced by MAGIC: extend the threshold of an IACT technique down to a ~20 GeV domain.

Technological novelties introduced by MAGIC:
a.  2 GHz sampling of the signal in a wide bandwidth >200 MHz;

b.   Parabolic tessellated reflector;
c.   Mat, hemispherical input window PMTs + tailored to these light guides for fast timing and higher detection efficiency;
d.   Analog signal transmission via optical fibers;
e.   Active Mirror Control system, for optimal PSF in the focus;
f.   Sealed, temperature controlled and stabilized imaging camera;
g.   Light-weight reflector frame made of reinforced carbon-fiber.

# 9. Fourth Generation Instruments

9.1 Cherenkov Telescope Array — The Major Instrument

The series of "Towards a Major Atmospheric Cherenkov Detector" workshops, taking place between 1989 in Crimea (seen historically the workshop number "0"), and 2005 in Paris (the last one), served its purpose. The researchers reached a consensus that one needed to unify the efforts of different collaborations and of the entire community and to move towards one major instrument.

In 2006 a new collaboration was formed for building the Cherenkov Telescope Array. This collaboration, which counts over ~1500 researchers, includes practically all the researchers worldwide working with the atmospheric Cherenkov technique and many newer groups with interest in exploring the sky in gamma rays with unprecedented sensitivity.

In the meantime, the CTA collaboration has produced advanced prototypes of its constituent telescopes and moved into the construction phase. Originally about 100 telescopes of 23 m, 12 m and 4-7 m size were planned for building in the southern and northern observatories, covering the energy range from 10 GeV to more than 100 TeV [69,70]. This is going to be the major ground-based instrument for conducting astrophysics by means of gamma rays for the next few decades.

The first Large Size Telescope (LST) is in its final stage of commissioning. It has already measured gamma-ray signal from dozens of sources. Many publications are scheduled from this telescope [71].

Similarly, the prototypes of the Middle Size Telescope (MST), the double mirror, 9.7 m diameter Schwarzschild-Couder (SCT) telescope in Arizona and the small size (SST), 4 m prototype ASTRII telescope in Sicily have been built and successfully commissioned, see [71] and the links therein.

One of the advances of the CTA telescopes can be considered as the wide field of views of its telescopes. The study of the wide-FoV prime-focus telescopes began in 2005 with publication [72]. Soon this has been expanded by the study of even wider FoV IACTs of a more complex design, including two optical elements [73,74]. The ASTRII and SCT followed the design [73].

The CTA is planning to operate LST, MST and SST telescopes of ~4.5°, ~8° and ~10° apertures, correspondingly. The technology of these novel, fourth generation telescopes has been refined practically everywhere. After saturating the gamma-ray detection efficiency of individual telescopes, the CTA is pursuing the plan to use a large number of such telescopes to cover a large area, for providing an exclusively high sensitivity.

Another technological progress of the CTA is the use of the advanced light sensors, such as the classical PMTs with strongly improved parameters as well as the so-called SiPM, see more on these below.

### 9.1.1. Enhanced Quantum Efficiency PMTs

In the first stage of the development work, initiated in ~2004, the researchers from MAGIC, cooperating with the companies Electron Tubes Enterprises (London), Photonis (France) and Hamamatsu (Japan), succeeded in increasing the peak QE of PMTs with bialkali photo cathode from stagnating over ~40 years value of 25–27% up to ~32–35%. Subsequently those PMTs were dubbed as "Superbialkali" type [75].

The second stage of the development was started by a group of researchers cooperating with Hamamatsu and Electron Tubes in the frame of the CTA collaboration in 2009. One of the novel technologies applied in the MST and LST telescopes of the CTA collaboration is the use of novel 1.5-inch size PMTs with significantly improved parameters. At the end of the development work the PMTs from Hamamatsu showed a somewhat better performance than those from Electron Tubes and thus were selected for the use in CTA.

It should be noted that those developments revised all aspects of a classical PMT, including, for example, the light emission of its dynode system, see Figure 7a. The latter effect caused a high-rate of after-pulses [76]. A dedicated re-design reduced that negative effect. The novel PMTs became commercially available from Hamamatsu in 2014. They show an average peak quantum efficiency of ~43%, electron collection efficiency on the first dynode of 94–98% (for wavelengths ≥ 400 nm) and an after-pulsing rate of ≤0.02% for the set threshold of ≥4 ph.e. The pulses from the 7-dynode PMTs measure ~2.5 ns Full Width at Half Maximum (FWHM) level. The achieved record parameters make this PMT currently the best in the world [77].

Use of such PMTs allows one to significantly lower the energy threshold of both the MST and LST telescopes. In the latter case a threshold of ~20 GeV has already been reported.

### 9.1.2. SiPM-Based Imaging Camera

The emerging at the end of the 1990s of a new semiconductor light sensor technology, dubbed as SiPM (Silicon Photo Multiplier), received a strong boost in the development, especially for the possible use in the MAGIC IACT and the EUSO space mission [78]. The parameters of the sensors started rapidly improving; already in 2008-2010 the majority of parameters were almost saturated in pilot productions, as for example, a peak photon detection efficiency (PDE) of ~60% along with a cross-talk of ~2.5% level has been reported [79].

One of the remaining issues that still exists, but to a less degree, is the problem of cross-talk, see Figure 7b [80,81].

Researchers started building the first custom-segments of imaging cameras [80]. The first full-scale SiPM-based camera has been built and installed on the left-over mechanical mount of the HEGRA third telescope in La Palma. Since then, the telescope dubbed as FACT was in successful operation, in recent years in a robotic regime [82]. The potential of these relatively old SiPMs could not be fully exploited mainly due to the above-mentioned crosstalk effect, which prevented from operating these close to the maximum available PDE.

With time the SiPM parameters have significantly improved and both the double mirror design prototype telescopes ASTRII SST and the 9.7 m Schwarzschild-Couder (SCT) next to the Veritas telescopes in Arizona, use SiPM-based cameras [71].

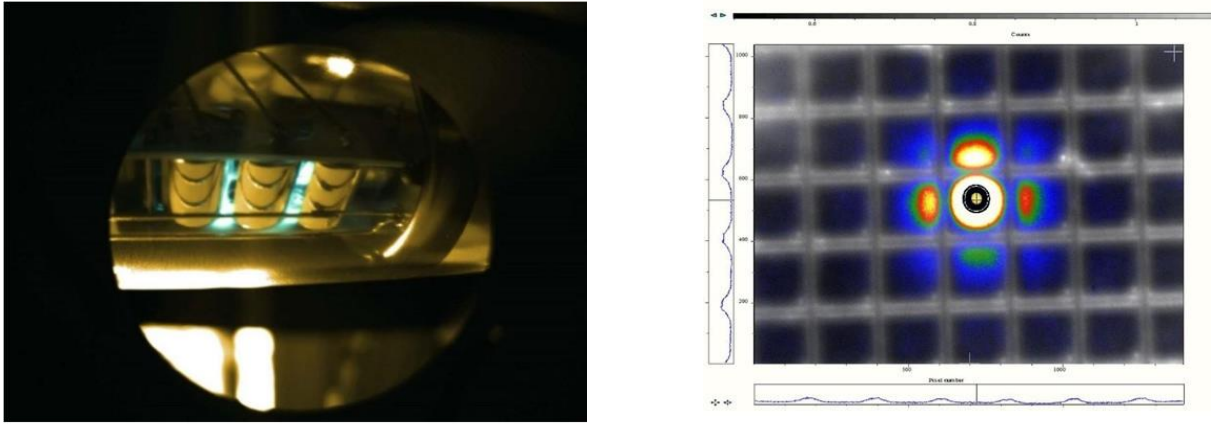

Figure 7. Left: photo shows the light emission leaking through the space between the dynode system of the Hamamatsu R8619 PMT. Part of this light arrives at the photocathode, causing high-level after-pulsing. Later, the company installed baffles, which reduced the negative effect, see [76] for details. Right: Light emission microscopy of a pilot SiPM sample (prod. By B. Dolgoshein and team) under operational voltage. The narrowly focused (<4 µm), weak laser beam shoots at the location of the yellow-black dot on the surface of a 100 µm size cell. One can see that also the neighbor cells emit light. This is the essence of the cross-talk effect; a single incident photon can fire more than one cell.

The size of the largest SiPM is limited to ≤10 mm. This limits their direct application mostly in SSTs. For using those in MSTs or LSTs, either one needs to use a significantly higher number of sensors and readout channels (compared to the number of current PMTs), or to group a large number of SiPMs for imitating one single larger-size sensor [83]. Please note that it does not make much sense to use sensors with a size much less than the optical PSF of a given telescope. Which of the above options will turn to be viable in future will depend on the cost evolution of those sensors and of the readout channels, as well as the ready availability of integral readout solutions.

9.2. TAIGA

An interesting hybrid approach has chosen the TAIGA pilot instrument in Tunka
valley near lake Baikal. One of the main goals is to explore the gamma-ray energy range from several to 100s of TeV. TAIGA includes 120 HiSCORE stations (improved version of the former timing array AIROBICC), deployed in an area of 1 km$^2$ [84], two 4 m class IACTs with a 9.6° FoV [85] and other types of Cherenkov light and particle detectors. The number of IACTs is planned to be completed to four until the end of the next year. A combination of the timing and imaging air shower detection techniques allows one observing the novel "hybrid stereo" mode: the core position and the incoming direction of a given shower can be obtained from HiSCORE, while its images from the IACTs will help measuring the type (gamma or hadron) and the energy.

Placed at a distance of about 600 m from each other, the four IACTs with HiSCORE will compose a sensitive detector with a collection area in excess of 1 km2.

The TAIGA approach has the promise to offer a cost-effective solution for building a highly sensitive detector of a very large area.

### 9.3. LHAASO

LHASSO is a multi-component, very large size cosmic and gamma-ray detector. It is located in Sichuan Province of China, at a mountain altitude 4410 m a.s.l. It plans to measure the cosmic and gamma rays in the energy range of $\geq 10^{12}$ eV and $10^{11}$–$10^{15}$ eV, respectively. LHAASO is designed to measure electrons, muons, Cherenkov and fluorescence light. Recently it made an important discovery, with a dozen so-called PeVatron sources identified [86]. The Wide Field-of-view Cherenkov Telescope Array (WFCTA) of LHAASO includes eighteen telescopes, based on reflectors of 5 m$^2$ area [87]. Composite SiPM pixel imaging cameras with a FoV of 16° × 16° are installed at their focal planes. Interestingly, the telescopes are portable, so their configuration and location can be easily changed.

# 10. Conclusions

It is remarkable to see the progress, made from the first detection of cosmic rays via Cherenkov light emission in the atmosphere in 1953 until present day. The original tiny Cherenkov telescope has served its purpose. After about 70 years the Veritas, MAGIC, H.E.S.S, and now the LST/CTA telescopes allow one to measure a significant gamma-ray signal from the Crab Nebula in less than a minute. Essentially, the speculative presentation of Cocconi from 1959 came true, if not exactly in the way he had predicted. Today we have
instruments with resolutions of 0.05–0.1°, which can measure a gamma-ray signal from the Crab Nebula with the signal to noise ratio of 300:1 for energies above 100 GeV. In the near future, the completed CTA instrument will further enhance that signal to noise ratio. The hundreds of new discoveries made at the very high energies established the firm place of the ground-based gamma-ray astrophysics as one of the rapidly evolving successful branches in astronomy. One can anticipate much more numerous, important results in cosmic rays, in multi-wavelength and messenger astrophysics and cosmology to become available within the next ~10 years.

Acknowledgments: The author wants to express his gratitude to Derek Strom for critically reading the manuscript.